\documentclass[12pt]{article}
\usepackage{amsmath}
\usepackage{enumerate}
\begin{document}
\def\be{\begin{equation}}
\def\bea{\begin{eqnarray}}
\def\ee{\end{equation}}
\def\eea{\end{eqnarray}}
\def\d{\partial}
\def\eps{\varepsilon}
\def\la{\lambda}
\def\b{\bigskip}
\def\nn{\nonumber \\}
\def\p{\partial}
\def\t{\tilde}
\def\h{{1\over 2}}
\newcommand{\comment}[2]{#2}

\makeatletter
\def\blfootnote{\xdef\@thefnmark{}\@footnotetext}  
\makeatother

\begin{center}
{\LARGE Towards higher dimensional black rings}
\\
\vspace{18mm}
{\bf   Yuri Chervonyi}
\vspace{14mm}

Department of Physics,\\ University at Albany (SUNY),\\ Albany, NY 12222, USA\\ 

\vskip 10 mm

\blfootnote{ichervonyi@albany.edu}

\end{center}

\begin{abstract}

Despite all attempts, exact solutions for black rings in more than five dimensions remain elusive. In this paper we clarify some of the reasons for that, in particular we show that a peculiar symmetry of the five--dimensional black ring - separability of the base - cannot occur in dimensions higher than five. We also construct supersymmetric solutions that have symmetries of 5D supersymmetric black ring and show that they do not have regular horizons.

\b

\end{abstract}

\newpage


\section{Introduction}

\noindent

Stationary, asymptotically flat four--dimensional black holes must have only spherical event horizons \cite{Hawking}. On the other hand, black holes in higher dimensions are less restricted and allowed to possess horizons of various topologies. The first example of a black hole with a non--spherical horizon was found by Emparan and Reall \cite{BRwick}, who used Kaluza--Klein C--metric of \cite{BRorigin} to construct five--dimensional black ring with the horizon of topology $S^1\times S^2$.
Alternative methods of constructing five--dimensional black rings are based on the generalized Weyl ansatz \cite{BRWeyl} or the inverse scattering method \cite{BelinskiZaharkov}. The latter approach was used to construct the black ring with two rotations \cite{BRinvsc} and several other configurations of black rings and black holes in five dimensions \cite{InvScatRings}. Unfortunately both the Weyl and the inverse scattering approaches rely on the presence of $D-2$ commuting Killing vectors, so they cannot be used to construct black rings  in $D>5$. In the absence of methods for finding the exact solutions with non--spherical topology in higher dimensions several approximate techniques have been developed, for example, the matching asymptotic expansions \cite{MatchingAsymp} and the blackfold effective theory \cite{Blackfolds}. Along with numerical \cite{NumReviews,NumNewTopologies} and approximate \cite{ApproxNewTopologies} methods, the blackfolds have been used to shows existence of solutions with non--spherical topologies such as helical black strings/rings, non--uniform black cylinder and several other possibilities \cite{NewTopologies,NewTopologies2}.
However, despite all recent results the exact solutions with non--spherical horizon topology are known only in  $D=5$. Approximate and numerical higher--dimensional solutions were constructed in \cite{ApproxhighBR, AppHighRingsOthers} and \cite{NumHighRings} respectively.

Another interesting direction towards finding black rings in higher dimensions is based on using supersymmetry. The supersymmetric five--dimensional black ring was constructed in \cite{SUSYring} and extended to a larger class of solutions in \cite{SUSYringTubes,SUSYrings}. However, analogously to the neutral case, SUSY black rings in higher dimensions ($D>5$) are still unknown, moreover there is even less progress in this direction. To summarize, the exact solutions for higher dimensional black rings remain elusive, and in this paper we clarify some of the reasons for that.

\hfill

This paper has the following structure.
In Section 2 we try to generalize a common symmetry of five--dimensional black holes and black rings to higher dimensions and show that solutions with such symmetries do not exist. In Section \ref{SecSUSYBR} we show that symmetries of 5D supersymmetric black ring do not survive in higher dimensions as well.

\section{Separability of the neutral black ring}
\label{SecNeutralBR}

The neutral  five--dimensional black ring was constructed in \cite{BRwick,BRWeyl} and reviewed in Appendix \ref{AppMPandBR}.
We  begin by noting that the black ring metric \eqref{AppBRmetric} has a structure of a $t$--fiber over the four--dimensional base, which is conformally separable:
\bea\label{BRbase}
ds_{base}^2=\frac{R^2 F(x)F^2(y)}{(x-y)^2}\left[-\Big(\frac{G(y)d\phi^2}{F^2(y)}+\frac{dy^2}{F(y)G(y)}\Big)+\Big(\frac{G(x)d\psi^2}{F^2(x)}+\frac{dx^2}{F(x)G(x)}\Big)\right].
\eea
Moreover it is separable in two different coordinate systems as we will show in a moment. In order to do that we recall that separability of the massless Hamilton--Jacobi equation
\bea\label{MasslessHJ}
g^{MN}\d_M S\d_N S=0
\eea
can be encoded in the conformal Killing tensor $\mathcal{K}$ of rank two\footnote{Killing tensor generalizes the well--known notion of Killing vector.}, which satisfies the following equation \cite{Tachibana}
\bea\label{CKTeq}
\nabla_{(M} \mathcal{K}_{NP)}=W_{(M}g_{NP)},
\eea
where $W_M$ is the associated vector. 
To find conformal Killing tensors one can solve the general equation \eqref{CKTeq} or extract it from the metric via \eqref{MasslessHJ} if it is written in the separable coordinates. Usually such tensors are used to construct the conserved quantities through
\bea
I=\mathcal{K}^{MN}\d_M S \d_N S,
\eea
or to extract the separable coordinates. Procedures of constructing Killing tensors from the metric, extracting separable coordinates from the tensors are described in Section 2 of \cite{KYTinSt} and here we outline the results.

The massless Hamilton--Jacobi equation \eqref{MasslessHJ} separates if there exists a function $f$, such that
\bea
g^{MN}=\frac{1}{f}\left(X^{MN}+Y^{MN} \right), \quad \d_x Y^{MN}=\d_y X^{MN}=0.
\eea
Then the conformal Killing tensor can be read off as
\bea\label{CKTfromSep}
\mathcal{K}^{MN}=X^{MN}.
\eea
Applying this procedure to the base of the black ring \eqref{BRbase} we get
\bea
(\mathcal{K}^{MN})^{(x)}\d_N\d_N=\frac{F^2(x)}{G(x)}\d_\psi^2+F(x)G(x)\d_x^2.
\eea
Here the index ${}^{(x)}$ indicates that this tensor was read off from the $x$--dependent part of the metric\footnote{Note that one always can read off another tensor from the $y$--dependent part of the metric, but these tensors will not be independent.}. Lowering the indices gives
\bea\label{CKTring1}
(\mathcal{K}_{MN})^{(x)}dx^Mdx^N=\left[\frac{R^2 F(x)F^2(y)}{(x-y)^2}\right]^2\left[\frac{G(x)d\psi^2}{F^2(x)}+\frac{dx^2}{F(x)G(x)}\right].
\eea
The solution \eqref{AppBRmetric},\eqref{BRbase} contains both the black ring and the black hole with one rotation (for details see Appendix \ref{AppMPandBR}), which allows us to make an assumption that higher dimensional black rings are in the same class of solutions as the Myers--Perry black holes with one rotation \cite{MyersPerry}. So we can study higher--dimensional neutral black holes and based on the results make conclusions about the corresponding rings.

The Myers--Perry black holes are reviewed in Appendix \ref{AppMPandBR}, and here we start with writing the tensor \eqref{CKTring1} in the standard Myers--Perry coordinates for the static case. Substituting the map \eqref{AppStaticMap} into \eqref{CKTring1} we get
\bea\label{CKT2}
(\mathcal{K}_{MN})^{(x)}dx^M dx^N=\frac{r^6 \cos^2\theta}{2R^2}d\psi^2-r^4\cos^4\theta d\psi^2+\frac{r^6 \cos^2\theta}{R^2}\left[d\ln\frac{\cos\theta}{r}\right]^2,\quad m=2R^2.
\eea

We have found the conformal Killing tensor associated with separability of the base in the ring--like coordinates, and now we want to find all separable coordinate systems for static neutral black holes followed by analysis of the cases with one rotation. Solving the conformal Killing tensor equation \eqref{CKTeq} on the base of the Tangherlini solution \cite{Tangherlini} with $D\ge 5$
\bea\label{SchwBase}
ds_{base}^2=\frac{dr^2}{1-\frac{m}{r^{D-3}}}+r^2 (d\theta^2+\sin^2\theta d\phi^2 +\cos^2\theta d\Omega_{D-4}^2),
\eea
we obtain the following non-trivial conformal Killing tensors\footnote{Note that here we do not write expression for the associated vectors $W^{(i)}$ entering right--hand side of the conformal Killing tensor equation \eqref{CKTeq} because they can be easily recovered from the corresponding conformal Killing tensors $\mathcal{K}^{(i)}$.}:
\bea\label{CKTsSchw}
\mathcal{K}_{MN}^{(1)}dx^Mdx^N&=&r^4{(p+q)^{4/(D-3)}}\cos^2\theta \Big[\left(d\ln \cos\theta-
\frac{2}{D-3}d\ln[p+q]\right)^2+d\Omega^2_{D-4}\Big]\nn
\nn
\mathcal{K}_{MN}^{(2)}dx^Mdx^N&=&r^4{(q-p)^{4/(D-3)}}\cos^2 \theta \Big[\left(d\ln \cos\theta+
\frac{2}{D-3}d\ln[p+q]\right)^2+d\Omega^2_{D-4}\Big],\nn
\mathcal{K}_{MN}^{(3)}dx^Mdx^N&=&\frac{r^{D-1}}{r^{D-3}-m}dr^2,\quad q=r^{(D-3)/2},\quad p=\sqrt{q^2-m}.
\eea
Now we identify the separable coordinate systems associated with these conformal Killing tensors. 

Using the prescription \eqref{CKTfromSep} we extract the obvious tensor from \eqref{SchwBase}
\bea
(\mathcal{K}^{MN})^{(r)}\d_M\d_N=\frac{r^{D-3}-m}{r^{D-5}}(\d_r)^2\quad \Rightarrow \quad \mathcal{K}^{(r)}=\mathcal{K}^{(3)}.
\eea
Here the index ${}^{(r)}$ indicates that this tensor is associated with the $r$ coordinate. We conclude that $\mathcal{K}^{(3)}$ is associated with separability in the standard Myers--Perry coordinates.

Next by comparing the expressions \eqref{CKT2} and \eqref{CKTsSchw} we find that the tensor responsible for separation in the ring--like coordinates is 
\bea
\mathcal{K}^{(x)}=\frac{1}{2m}\left(\mathcal{K}^{(1)}+\mathcal{K}^{(2)}\right)+\mathcal{K}^{(3)}.
\eea
Thus we have associated two independent conformal Killing tensors with separable coordinate systems, namely the ring--like and standard Myers--Perry coordinates, but the complete solution of the conformal Killing equation \eqref{CKTeq} gives three independent tensors, so it is natural to ask about the meaning of the third one. To answer this question we recall the procedure of extracting coordinates from Killing tensors described in Section 2 of \cite{KYTinSt}. It was shown that the separable coordinates can be obtained from the eigenvectors of corresponding tensors. For the tensors \eqref{CKTsSchw} we find
\bea\label{CoordsTable}
\begin{array}{|r|c|c|c|}
\hline
      & \mathcal{K}^{(1)} & \mathcal{K}^{(2)} & \mathcal{K}^{(3)} \\ \hline
    \hat{x} & \displaystyle\frac{\sin\theta}{\sqrt{q+p}} & \sqrt{q+p}\sin\theta & q \\ \hline
    \hat{y} & -\displaystyle\frac{\cos\theta}{\sqrt{q+p}} & -\sqrt{q+p}\cos\theta & \theta \\ \hline
\end{array}
\qquad q=r^{(D-3)/2},\quad p=\sqrt{q^2-m}.
\eea
To summarize we see that the expressions for conformal Killing tensors \eqref{CKTsSchw} have universal character in all dimensions. It means that the bases of \textit{neutral static black holes have the same symmetries in any dimension} and five dimensions is not an exception. Now we will check this statement in the rotating case. Direct solving the conformal Killing tensor equation and treating rotation as a perturbative parameter shows that introduction of rotation decreases the number of conformal Killing tensors down to one in dimensions higher than five. In five dimensions turning on rotation destroys one of the tensors, and the survivors are $\mathcal{K}^{(3)}$ and $\mathcal{K}^{(1)}+\mathcal{K}^{(2)}$ corresponding to separation in the standard Myers--Perry and ring--like coordinates respectively. In dimensions higher than five only one tensor responsible for separation in the Myers--Perry coordinates survives. We conclude that if higher dimensional black rings are in the same class of solution as black holes they would not have the separable bases unlike their five--dimensional counterpart.


\section{Supersymmetric black rings}
\label{SecSUSYBR}

In the previous section we have been focusing on the neutral black rings and now we switch to the analysis of their supersymmetric counterparts. Consider the five dimensional SUSY black ring constructed in \cite{SUSYring} and recall that it was found 
by utilizing a very special feature of the neutral black ring - separability of the fiber. The solution was written in the form
\bea\label{SUSYBR5D}
ds^2=-f^2(dt+\omega)^2+f^{-1}h_{mn}dx^mdx^n,\quad m,n=1,..,4,
\eea
and the fiber one--form $\omega$ was assumed to satisfy
\bea
\omega=\omega_\phi d\phi+\omega_\psi d\psi,\quad \d_x\omega_{\psi}=\frac{y^2-1}{1-x^2}\d_y\omega_{\phi}.
\eea
Furthermore 5D SUSY ring was embedded into M theory in \cite{SUSYringTubes}, which is a good starting point for constructing higher dimensional SUSY rings. Following \cite{SUSYringTubes} the solution reads
\bea\label{11sol}
ds_{\it 11}^2 &=&  ds^2_{\it 5} + X^1 \left( dz_1^2 +
dz_2^2 \right) + X^2 \left( dz_3^2 + dz_4^2 \right) +
X^3 \left( dz_5^2 + dz_6^2 \right),\nonumber \\
\mathcal{A} &=& A^1 \wedge dz_1 \wedge dz_2 + A^2 \wedge dz_3 \wedge dz_4 +
A^3 \wedge dz_5 \wedge dz_6.
\eea
Here $\mathcal{A}$ is the three-form potential with four-form field
strength $\mathcal{G}=d\mathcal{A}$. The solution is specified by three scalars $X^i$, and three one-forms
$A^i$, which are defined on a five-dimensional
spacetime with metric $ds^2_{\it 5}$:
\bea
\label{5dsol}
ds_{\it 5}^2 &=& -(H_1 H_2 H_3)^{-2/3} (dt + \omega)^2 +
(H_1 H_2 H_3)^{1/3} d{\bf x}_{\it 4}^2,\nonumber \\
A^i &=& H_i^{-1}(dt +\omega) +
 \frac{q_i R^2}{r^2+R^2\cos^2\theta} (\sin^2\theta d\psi -\cos^2\theta d\phi ),\\
X^i &=& H_i^{-1} (H_1 H_2 H_3)^{1/3},\nn
d{\bf x}_{\it 4}^2&=&(r^2+R^2\cos^2\theta)\left(\frac{dr^2}{r^2+R^2}+d\theta^2 \right)+(r^2+R^2)\sin^2\theta d\psi^2+r^2 \cos^2\theta d\phi^2.\nonumber
\eea
Here $H_i$ are harmonic functions on the flat four--dimensional base $d\mathbf{x}_4^2$. 

Now we want to write the prototype of 7D SUSY black ring while keeping the symmetries of \eqref{11sol}, such as the flat base and several $Z_2$ symmetries. First we extend the flat base to six dimensions effectively absorbing $z_5,z_6$ into $d\mathbf{x}_4^2$ and focus on symmetries associated with the rest of $z_i$. Recalling the equations of motion in 11D SUGRA
\bea\label{EOM11D}
R_{MN}-\frac{1}{12}\left(\mathcal{G}_{MABC}\mathcal{G}_N{}^{ABC}-\frac{1}{12}g_{MN}\mathcal{G}^2 \right)=0,\nn
d\star\mathcal{G}+\frac{1}{2}\mathcal{G}\wedge \mathcal{G}=0
\eea 
we note that even though, for example, $(z_1,z_2)\to -(z_1,z_2)$ is not a symmetry of the field strength $\mathcal{G}$, it is a symmetry of equations of motion \eqref{EOM11D}. The rest of the symmetries together with their restrictions on the metric and three--form are collected in the following table\footnote{Here $z_i$ are denoted as $i$. For example, $g_{z_1z_1}=g_{11}$, etc.}:
\bea\label{SymmsTable}
\begin{array}{|l|l|}
\hline
     \mbox{Symmetry} & \mbox{Prohibited~expressions} \\ \hline
     (z_1,z_2)\to -(z_1, z_2) & g_{13}, g_{14}, g_{23}, g_{24}\\ 
      & \mathcal{A}_{13}, \mathcal{A}_{14}, \mathcal{A}_{23}, \mathcal{A}_{24}\\ \hline
     z_1\leftrightarrow z_2, z_3\leftrightarrow z_4& g_{11}\ne g_{22}, g_{33}\ne g_{44}\\ \hline
     (z_1,z_3)\to -(z_1, z_3) & g_{12}, g_{34}\\ \hline
\end{array}
\eea
The first line constrains the three form $\mathcal{A}$ and partially fixes the metric.
The rest constrains the metric resulting in the following ansatz
\bea\label{11Dansatz}
ds_{11}^2&=&-\hat{H}_1(dt+\omega)^2+\hat{H}_2 d\mathbf{x}_6^2+\hat{H}_3(dz_1^2+dz_2^2)+\hat{H}_4(dz_3^2+dz_4^2),\nn
\mathcal{A}&=&A^1\wedge dz_1\wedge dz_2+A^2\wedge dz_3\wedge dz_4+\mathcal{C},
\eea
where $\hat{H}_{i}, i=1,..,4$ are unknown functions, $d\mathbf{x}_6^2$ is the flat six--dimensional space, $A^{1,2}, \mathcal{C}$ are respectively one-- and three--forms on the seven--dimensional base. Performing the dimensional reduction along one of $z_i$ followed by three T dualities along remaining $z_i$ and finally S duality gives the solution in IIB SUGRA:
\bea
ds_{10}^2&=&-\tilde{H}_1^2(dt+\omega)^2+\tilde{H}_2^2(dz_2+fdt+\alpha)^2+\tilde{H}_3^2 d\mathbf{x}_6^2+\tilde{H}_4^2(dz_{3}^2+dz_4^2),\nn
B&=&\beta_1dt+\beta_2dz_2+gdtdz_2+\tilde{\omega}_2, \quad e^{2\Phi}=\tilde{H}_5.
\eea
Here one--forms  $\omega,\alpha$ are defined on the flat six--dimensional base. Solving the Killing spinor equations for this ansatz reveals that the most general solution is governed by the chiral null model \cite{ChNullModel}:
\bea
ds^2&=&\frac{2}{H}dz_2(dt+\omega)+{\cal F}dz_2^2+d\mathbf{x}_6^2+(dz_3^2+dz_4^2),
\quad e^{-2\phi}=H,\nn
d\star_6(dH)&=&0,\quad d\star_6(d\omega)=0,\quad d\star_6(d[H {\cal F} ])=0.
\eea
This family of solutions does not admit a horizon with a non--zero area, but does give rise to a stretched horizon \cite{Sen} and upon dualization to D1--D5 frame may lead to completely regular geometries \cite{RegD1D5}.

We conclude that the natural extension 
of 5D SUSY black ring to higher dimensions by keeping its symmetries (the flat base and several $Z_2$ symmetries) leads to the chiral null model, which means the absence of 
a horizon. So in order to produce the finite ring--like horizons in higher dimensions 
one must consider the non--flat bases.


\section{Conclusions}

In the first part of this work we have shown that unlike neutral static black holes, which have the same symmetries in all dimensions, rotating higher dimensional black holes/rings do not. In particular, we found that separability of the base in more than one coordinate system is a special feature of the low--dimensional rotating black holes ($D\le 5$), which makes the 5D black ring metric to have such a simple structure. Our results show that if higher dimensional black rings are described by the same class of solutions as the black holes (as it occurs in 5D), then their metric will not have a simple and symmetric structure.

In the second part of this work we show that generalization of
5D supersymmetric black ring to higher dimensions while keeping its symmetries (in particular, the flat base) results in vanishing horizon. It would be interesting to extend this analysis to the non--flat bases.


\section*{Acknowledgments}

I am very grateful to my advisor Oleg Lunin for many discussions and for his guidance throughout this work. This work was supported by NSF grant PHY-1316184.


\appendix

\section{Myers--Perry black holes and the neutral black ring}
\label{AppMPandBR}
\renewcommand{\theequation}{A.\arabic{equation}}
\setcounter{equation}{0}

In this appendix we review neutral rotating black holes and the neutral five--dimensional black ring, in particular we are interested in black holes/ring with one rotation. Starting with the Myers--Perry black hole \cite{MyersPerry}, setting all the rotation parameters except one to zero and introducing 
\be
\mu_1=\sin \theta\equiv s_\theta,\quad\cos\theta\equiv c_\theta
\ee
one gets
\bea\label{AppMPmetric}
ds^2&=&-dt^2+\frac{m}{(r^2+a^2c_\theta^2)r^{D-5}}\Big(dt+a s_\theta^2 d\phi\Big)^2+(r^2+a^2c_\theta^2)\left(\frac{dr^2}{r^2+a^2-mr^{5-D}}
+d\theta^2\right)\nn
&&\quad+(r^2+a^2)s_\theta^2 d\phi^2+r^2c_\theta^2d\Omega_{D-4}.
\eea
Note that even though in this paper we study separability of the bases, we should mention that the whole Myers--Perry metric in any dimensions is also separable \cite{Kub,KYTinSt}.

Next we recall the neutral black ring with one rotation \cite{BRElvang}\footnote{Note that here comparing to \cite{BRElvang} we swapped $\phi$ and $\psi$ to be consistent with black holes.}
\bea\label{AppBRmetric}
ds^2 &=& -\frac{F(x)}{F(y)} \left(dt+ R\sqrt{\lambda\nu} (1 + y) d\phi\right)^2  \\
&&+\frac{R^2F(x)F(y)^2}{(x-y)^2}\left[ -\frac{1}{F(y)^2} \left( G(y) d\phi^2 + \frac{F(y)}{G(y)} dy^2 \right)+ \frac{1}{F(x)} \left( \frac{dx^2}{G(x)} + \frac{G(x)}{F(x)}d\psi^2\right)\right],\nonumber
\eea
where
\bea\label{AppFGdef}
F(\xi) = 1 - \lambda\xi, \quad  G(\xi) = (1 - \xi^2)(1-\nu \xi),\quad -1\le x\le 1,\quad -\infty<y\le-1.
\eea
In order to avoid conical singularity at $x=-1$ and $y=-1\ne x$ one must set \cite{BRwick}
\bea
\Delta\phi=\Delta\psi=
\frac{2\pi\sqrt{1+\lambda}}{1+\nu}.
\eea
Further one needs to avoid a singularity at $x=1$ which can be done in two ways
\bea\label{Applambdadef}
\lambda_r=\frac{2\nu}{1+\nu^2},\quad \lambda_h=1,
\eea
where the first choice corresponds to a black ring and the second one to a black hole.
In particular the metric \eqref{AppBRmetric} with $\lambda=1$ and the five--dimensional neutral black hole with one rotation (\eqref{AppMPmetric} with $D=5$) are related through
\bea\label{AppMP-ringMap}
r^2&=&\frac{2R^2(1-y)(1-\nu x)}{x-y},\quad c_\theta^2=\frac{(1+x)(1-y)}{2(x-y)},\nn
m&=&2R^2-2\nu R^2+a^2, \quad a=2\sqrt{\nu}R,\nn
t_{MP}&=&\frac{t_{BR}}{\sqrt{\alpha}},\quad \phi_{MP}=\frac{\phi_{BR}}{\sqrt{\alpha}}, \quad \psi_{MP}=\frac{\psi_{BR}}{\sqrt{\alpha}},\quad \alpha=\frac{2}{(1+\nu)^2}.
\eea
For the static black hole $\lambda=1,\nu=0$ we get
\bea\label{AppStaticMap}
F(\xi)=1-\xi,\quad G(\xi)=1-\xi^2,\quad x=-1+\frac{4R^2c_\theta^2}{r^2},\quad y=-1-\frac{4R^2s_\theta^2}{r^2-2R^2},\quad m=2R^2.\nn
\eea
Finally the flat space limit is obtained by setting $\nu=0$ followed by writing
\bea
x=-1+4R^2\tilde{x},\quad y=-1+4R^2\tilde{y}
\eea
and sending $R$ to zero
\bea
r^2&=&\frac{1}{\tilde{x}-\tilde{y}},\quad c_\theta^2=\frac{\tilde{x}}{\tilde{x}-\tilde{y}},\quad s_\theta^2=\frac{\tilde{y}}{\tilde{y}-\tilde{x}}.
\eea


\end{document}